\documentclass[oldversion]{aa}  

\usepackage{graphicx}
\usepackage{txfonts}
\usepackage{hyperref}
\usepackage{xcolor}
\newcommand{\hi}{{\sc H\,i}}
\newcommand{\HIbf}{\mbox{H\hspace{0.155 em}{\footnotesize \bf I}}}
\newcommand{\HIsl}{\mbox{H\hspace{0.155 em}{\footnotesize \sl I}}}
\newcommand{\mhi}{$M_\mathrm{HI}$}

\newcommand{\mhtwo}{$M_\mathrm{H_2}$}

\newcommand{\mstar}{$M_\star$}

\newcommand{\mJybeam}{mJy beam$^{-1}$}
\newcommand{\msun}{{M$_\odot$}}

\newcommand{\kms}{$\,$km$\,$s$^{-1}$}
\newcommand{\ltsima} {$\; \buildrel < \over \sim \;$}
\newcommand{\gtsima} {$\; \buildrel > \over \sim \;$}
\newcommand{\lta} {\lower.5ex\hbox{\ltsima}}
\newcommand{\gta} {\lower.5ex\hbox{\gtsima}}

\newcommand{\Nii}{[{\sc N$\,$ii}]}

\begin{document}

   \title{Neutral hydrogen gas within and around NGC~1316}

   \author{
   P.~Serra \inst{1}\fnmsep\thanks{paolo.serra@inaf.it},
   F.~M.~Maccagni \inst{1},
   D.~Kleiner \inst{1},
   W.~J.~G.~de~Blok \inst{2,3,4},
   J.~H.~van~Gorkom \inst{5},
   B.~Hugo \inst{6,7},
   E.~Iodice \inst{8},
   G.~I.~G.~J\'{o}zsa \inst{6,7,9},
   P.~Kamphuis \inst{10},
   R.~Kraan-Korteweg \inst{4},
   A.~Loni \inst{1,11},
   S.~Makhathini \inst{6,7},
   D.~Moln\'{a}r \inst{1},
   T.~Oosterloo \inst{2,3},
   R.~Peletier \inst{3},
   A.~Ramaila \inst{6,7},
   M.~Ramatsoku \inst{1},
   O.~Smirnov \inst{6,7},
   M.~Smith \inst{12},
   M.~Spavone \inst{8},
   K.~Thorat \inst{6,7},
   S.~C.~Trager \inst{3},
   \and
   A.~Venhola \inst{13}
          }
          
   \authorrunning{Serra et al.}

   \institute{
   INAF - Osservatorio Astronomico di Cagliari, Via della Scienza 5, I-09047 Selargius (CA), Italy
   \and
   Netherlands Institute for Radio Astronomy (ASTRON), Oude Hoogeveensedijk 4, 7991 PD Dwingeloo, the Netherlands
   \and
   Kapteyn Astronomical Institute, University of Groningen, PO Box 800, NL-9700 AV Groningen, the Netherlands
   \and
   Department of Astronomy, University of Cape Town, Private Bag X3, Rondebosch 7701, South Africa
   \and
   Department of Astronomy, Columbia University, New York, NY 10027, USA
   \and
   South African Radio Astronomy Oberservatory, Black River Park, 2 Fir Street, Observatory, Cape Town, 7925, South Africa
   \and
   Department of Physics and Electronics, Rhodes University, PO Box 94, Makhanda 6140, South Africa
   \and
   INAF - Astronomical Observatory of Capodimonte, Salita Moiariello 16, 80131, Naples, Italy
   \and
   Argelander-Institut f\"{u}r Astronomie, Auf dem H\"{u}gel 71, D-53121 Bonn, Germany
   \and
   Astronomisches Institut, Ruhr-Universit\"{a}t Bochum, Universit\"{a}tsstrasse 150, 44801 Bochum, Germany
   \and
   Dipartimento di Fisica, Universit\`{a} di Cagliari, Cittadella Universitaria, 09042 Monserrato, Italy
   \and
   School of Physics and Astronomy, Cardiff University, Queens Buildings, The Parade, Cardiff CF24 3AA, UK
   \and
   Astronomy Research Unit, University of Oulu, FI-90014 Oulu, Finland
   \\
             }

   \date{Received June 16, 2019; accepted July 15, 2019}

  \abstract
  {We present MeerKAT observations of neutral hydrogen gas (\hi) in the nearby merger remnant NGC~1316 (Fornax~A), the brightest member of a galaxy group which is falling into the Fornax cluster. We find \hi\ on a variety of scales, from the galaxy centre to its large-scale environment. For the first time  we detect \hi\ at large radii  (70 -- 150 kpc in projection), mostly distributed on two long tails associated with the galaxy. Gas in the tails dominates the \hi\ mass of NGC~1316:   $7\times 10^8$ \msun\ --- 14 times more than in previous observations. The total \hi\ mass is comparable to the amount of neutral gas found inside the stellar body, mostly in molecular form. The \hi\ tails are associated with faint optical tidal features thought to be the remnant of a galaxy merger occurred a few billion years ago. They demonstrate that the merger was gas-rich. During the merger, tidal forces  pulled some gas and stars out to large radii, where we now detect them in the form of optical tails and, thanks to our new data, \hi\ tails; while torques caused the remaining gas to flow towards the centre of the remnant, where it was converted into molecular gas and fuelled the starburst revealed by the galaxy's stellar populations.  Several of the observed properties of NGC~1316 can be reproduced by a $\sim$ 10:1 merger between a dominant, gas-poor early-type galaxy and a smaller, gas-rich spiral occurred 1 -- 3 Gyr ago, likely followed by subsequent accretion of satellite galaxies.}
   
   \keywords{
   Galaxies -- Galaxies: interactions -- Galaxies: ISM -- Galaxies: individual: NGC~1316
               }

   \maketitle

\section{Introduction}
\label{sec:intro}

NGC~1316 (Fornax~A) is a nearby, giant early-type galaxy and a well known merger remnant\footnote{Throughout this paper we assume a luminosity distance of 20 Mpc for NGC~1316, consistent with the recent estimates $(20.8 \pm 0.5_\mathrm{stat} \pm 1.5_\mathrm{syst})$ Mpc and $(18.8 \pm 0.3_\mathrm{stat} \pm 0.5_\mathrm{syst})$ Mpc by \cite{cantiello2013} and \cite{hatt2018}, respectively. We assume the same distance for all other neighbouring galaxies. At this distance, 1\arcmin\ corresponds to 5.8 kpc. We scale all quantities from the literature to our adopted distance.}. It is the brightest member of a galaxy group at the outskirts of the Fornax  cluster and, in fact, the brightest galaxy in the entire cluster volume \citep{schweizer1980,iodice2017}. The stellar body of NGC~1316 is characterised by numerous tails and loops \citep{schweizer1980,iodice2017} whose brightness and size imply at least one relatively major, recent merger with a mass ratio of approximately 10:1 or below \citep{schweizer1980,mackie1998,goudfrooij2001}. The merger(s) must have happened between $\sim1$ and $\sim3$ Gyr ago based on the expansion time of the outer stellar loops \citep{schweizer1980} and the spread in the ages of the globular clusters \citep{goudfrooij2001,sesto2016,sesto2018}, after which NGC~1316 has likely continued accreting smaller satellite galaxies \citep{iodice2017}.

Although NGC~1316's merger origin is unquestionable, not all pieces of the puzzle have fallen into place yet. In particular, the composition of the galaxy's interstellar medium remains difficult to understand. Given its stellar mass of $(6 \pm 2) \times 10^{11}$ \msun\ \citep{iodice2017} NGC~1316 hosts an anomalous amount of dust: $\sim2\times10^7$ \msun\  \citep{draine2007,lanz2010,galametz2012}  compared to the $\lesssim 10^6$ \msun\ expected based on known scaling relations for early-type galaxies \citep{temi2009,smith2012}. This could be explained if NGC~1316 formed and acquired its dust via a $\sim$10:1 merger between a dominant, dust-poor early-type galaxy and a smaller, dust-rich spiral. Indeed, \cite{lanz2010} estimated that a merging spiral with \mstar\ $=1\ \text{--}\ 6 \times 10^{10}$ \msun\ would bring a sufficient mass of dust to explain the observations. The problem is that the spiral would bring into the remnant a substantial amount of cold atomic and molecular gas, too: $2\ \text{--}\ 4\times 10^9$ \msun\ in total, which is not observed.

The most complete study of the cold gas content of NGC~1316 is \cite{horellou2001}. These authors detected $\sim6 \times 10^8$ \msun\ of  molecular hydrogen (H$_2$)  in the central region of the galaxy, and a further $\sim 5 \times 10^7$ \msun\ of  atomic hydrogen (\hi)  scattered in a few clouds at the outskirts of the stellar body, at a projected radius of $\sim30$ kpc. They did not detect any \hi\ in the centre of the galaxy down to $\sim10^8$ \msun\ and they did not detect any \hi\ at larger  radii. The latter is puzzling because the long ($\gtrsim100$ kpc) stellar tidal tails of NGC~1316 have most likely formed from the kinematically-cold stellar disc of the spiral progenitor involved in the merger. The spiral disc is gas-rich and, therefore, it would be reasonable to find gas at about the same radius (if not exactly at the same location)  as  the optical tails.

This is indeed the case in several late-stage mergers or merger remnants involving at least one gas-rich progenitor. Once the central stellar body has relaxed into an early-type morphology, little \hi\ is usually found in the inner regions while \hi\ is found at large radii, where stellar tails and loops are still visible. Such systems include, for example, NGC~3921 \citep{hibbard1996}, NGC~5128 \citep{schiminovich1994}, NGC~0680 and NGC~5557 \citep{duc2011},  NGC~7252 \citep{hibbard1994}, Arp~220 \citep{hibbard2000} and Mrk~315 \citep{simkin2001}. The lack of \hi\ at large radii in NGC~1316 is therefore a problem within the context of its supposed $\sim$10:1 early-type+spiral merger formation.

In this paper we present MeerKAT 21-cm \hi\ line observations that shed new light on the gas content of NGC~1316 and solve the above puzzle. We show that \hi\ is actually present at large radii around NGC~1316; that it coincides at least partially with some of the known optical tidal features; and that the distribution and total mass of cold gas  within and around NGC~1316 is in agreement with those expected for a $\sim$10:1 merger between a dominant early-type galaxy and a smaller, gas-rich spiral.

\section{MeerKAT observations and data reduction}

We observed NGC~1316 with MeerKAT  \citep{camilo2018}  in June 2018 as part of the telescope commissioning (Table \ref{table:obs}). The observation was carried out with 36 antennas (actually 40, but 4 did not deliver useful data). The baseline length varies between 29 m and 7.5 km. Of the 630 baselines, 15 are shorter than 100 m, 170 shorter than 500 m and 274 shorter than 1 km. We used the SKARAB correlator in the 4k mode, where 4096 channels sample the frequency interval 856 to 1712 MHz in full polarisation. The channel width is 209 kHz, corresponding to 44.5 \kms\ for \hi\ at redshift $z= 0$. The visibilities were recorded every 8 seconds. The total  integration time on NGC~1316  was 8 hours. We observed the bandpass and flux calibrator PKS~1934-638 for 10 minutes every 2 hours, and the gain calibrator PKS~0032-403 for 2 minutes every 12 minutes.


   \begin{table}
   {\centering
      \caption[]{MeerKAT observations and data products}
         \label{table:obs}
         \begin{tabular}{ll}
            \noalign{\smallskip}
            \hline
            \hline
            \noalign{\smallskip}
            Date & 2 June 2018 \\
            Observation ID & 1527936442 \\
            Pointing centre (J2000)                & RA 03h 22m 41.7s     \\
                                                    & Dec $-$37d 12m 30s     \\
            Total frequency range & 856 -- 1712 MHz \\
            Processed frequency range & 1381 -- 1500 MHz \\
            Channel width & 209 kHz \\
             & 44.5 \kms\ for \hi\ at $z=0$ \\
            Number of antennas & 36 \\
            Bandpass/flux calibrator & PKS~1934-638 \\
            Gain calibrator & PKS~0332-403  \\
            Time on target & 8 h \\
            \hi\ cube weighting &  Briggs \it robust~$=0.5$ \rm \\
             & 20\arcsec\ tapering \\
            \hi\ cube r.m.s. noise & 0.19 \mJybeam\ (cube centre) \\
             \hi\ restoring PSF & 36.7\arcsec\ $\times$ 28.1\arcsec FWHM \\
                & PA = 109 deg \\
            $3\sigma$ \hi\ column density & $2.7\times 10^{19}$ cm$^{-2}$ \\
            & (cube centre, single channel)\\
            Primary beam (FWHM)                         & 0.9 deg at 1.4 GHz     \\
            \noalign{\smallskip}
            \hline
            \noalign{\smallskip}
         \end{tabular}}
   \end{table}

For the purpose of this work we only processed  parallel hand  visibilities in the frequency interval 1381 -- 1500 MHz. We reduced the data with  a pipeline\footnote{https://meerkathi.readthedocs.io} currently under development at SARAO and INAF. The pipeline is based on \texttt{Stimela}\footnote{https://github.com/SpheMakh/Stimela}, a \texttt{Python}-based scripting framework which allows users to run several open-source radio interferometry software packages in a same script  using containers. The pipeline is flexible and allows users to define their own preferred data reduction strategy. For the purpose of this work we: \it i) \rm flagged the calibrators data with \texttt{AOFlagger} \citep{offringa2012} based on the Stokes Q visibilities; \it ii) \rm derived a time-independent, antenna-based, complex bandpass and flux scale with \texttt{CASA/bandpass} and \texttt{CASA/gaincal}; \it iii) \rm determined frequency-independent, time-dependent, antenna-based complex gains with \texttt{CASA/gaincal}; \it iv) \rm scaled the gain amplitudes to bootstrap the flux scale with \texttt{CASA/gaincal}; \it v) \rm applied bandpass and gains to the target visibilities with \texttt{CASA/applycal}; \it vi) \rm flagged the target data with \texttt{AOFlagger} based on the Stokes Q visibilities; \it vii) \rm iteratively imaged and self-calibrated the target radio continuum with \texttt{WSclean} \citep{offringa2014,offringa2017} and \texttt{MeqTrees} \citep{noordam2010}, respectively, solving for frequency-independent gain phase only with a solution interval of 2 minutes; \it viii) \rm subtracted the best continuum model from the visibilities; \it ix) \rm removed any residual continuum emission by fitting and subtracting a $3^\mathrm{rd}$ order polynomial to real and imaginary visibility spectra with \texttt{CASA/mstransform}; \it x) \rm Doppler corrected the visibilities to the barycentric reference frame with \texttt{CASA/mstransform}; \it xi) \rm iteratively imaged the target \hi\ emission with \texttt{WSclean} and created 3D clean masks with \texttt{SoFiA} (\citealt{serra2015a}; the \hi\ cube was made using Briggs \it robust~$=0.5$ \rm and 20\arcsec\ tapering in order to obtain a more Gaussian PSF main lobe compared to natural weighting while reaching a similar surface brightness sensitivity); \it xii) \rm created a 3D \hi\ detection mask, a total \hi\ image and an \hi\ velocity field with \texttt{SoFiA}, which we then validated with visual inspection; and \it xiii) \rm used the transformations in \cite{meyer2017} to convert the \hi\ image into \hi\ column density units, and the integrated \hi\ flux of each detection into an \hi\ mass assuming a distance of 20 Mpc.  Throughout the paper we adopt the optical definition of radial velocity.

The size of the Gaussian restoring PSF of the \hi\ cube is 36.7\arcsec\ $\times$ 28.1\arcsec\ at FWHM and the  position angle  is 109 deg  (north through east) . The cube noise is 0.19 \mJybeam\ per 44.5-\kms-wide channel.  This corresponds to a $3\sigma$ \hi\ column density sensitivity of $2.7\times 10^{19}$ cm$^{-2}$ in a single channel. Assuming a 100 \kms\ line-width, the $5\sigma$ point-source \mhi\ sensitivity is $5.9\times 10^6$ \msun\ at the assumed distance of 20 Mpc. 

Compared with the Very Large Array \hi\ cube of \cite{horellou2001}, ours has almost exactly the same velocity resolution (44.5 \it vs. \rm 41.7 \kms), $2\times$ smaller restoring PSF area ($0.33$ \it vs. \rm 0.67 arcmin$^2$), $3\times$ lower noise at the centre of the field (0.19 \it vs. \rm 0.6 \mJybeam), and $\sim2\times$ larger field of view diameter at FWHM of the primary beam ($0.9$ \it vs. \rm 0.5 deg).

\begin{figure*}
\centering
\includegraphics[width=18cm]{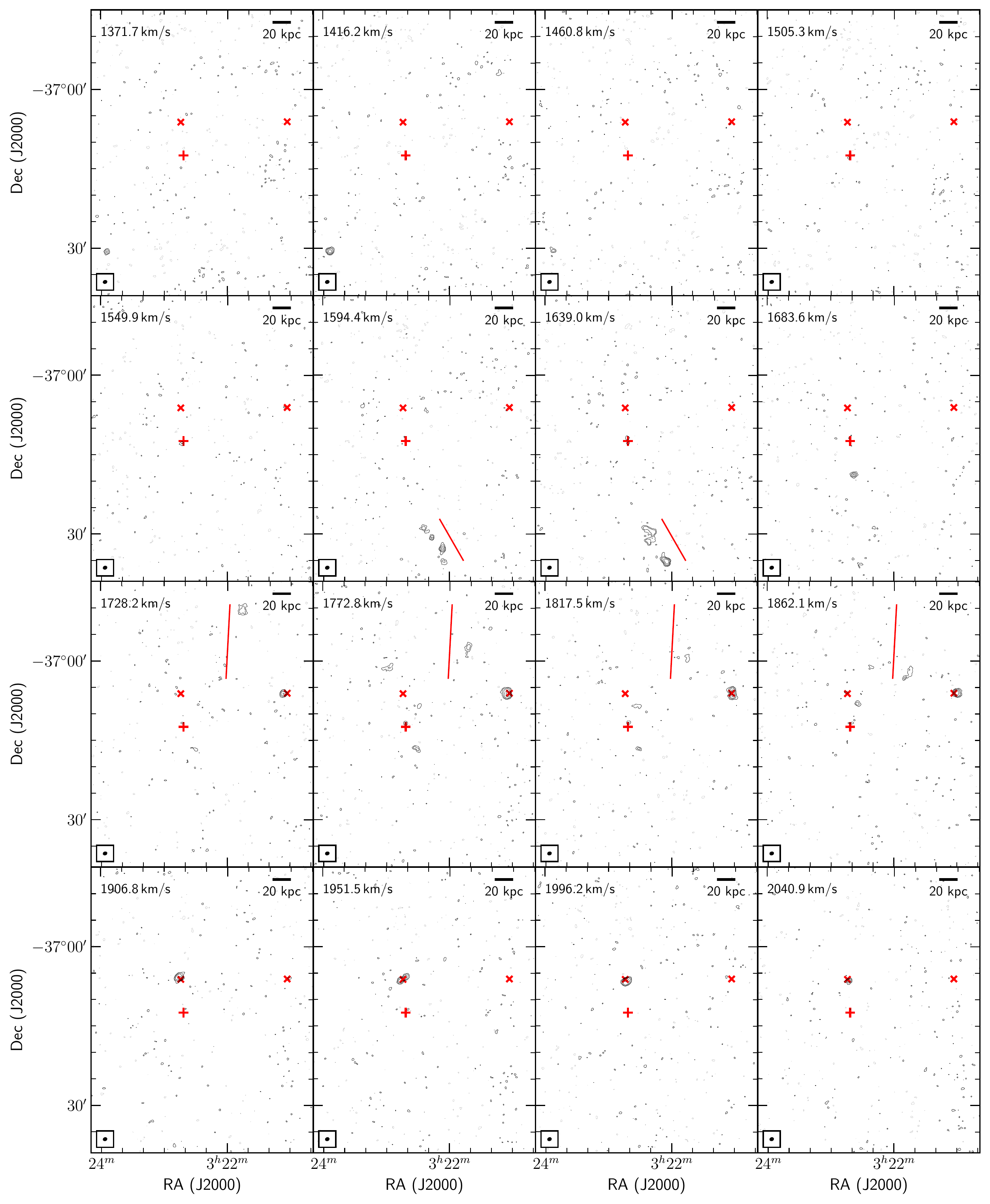}
\caption{Selected consecutive channel maps extracted from the \hi\ data cube of the NGC~1316 region obtained from the MeerKAT data. The figure shows channels in the recessional velocity range where we detect \hi. We list the velocity of each channel in the top-left corner. Velocities are in the barycentric reference frame (optical definition). The contour levels are $-0.5$ \mJybeam\ (grey dotted) and $0.5, 1, 2, 4, 8$ \mJybeam\ (black solid; for reference, the noise is 0.19 \mJybeam). The Gaussian restoring PSF in the bottom-left corner has a FWHM of 36.7\arcsec\ $\times$ 28.1\arcsec\ and position angle 109 deg. The scale bar in the top-right corner represents 20 kpc ($\sim3.5$\arcmin\ at a distance of 20 Mpc). The red ``+'' marker represents the centre of NGC~1316. The two red ``x'' markers represent NGC~1317 (east) and NGC~1310 (west), respectively. The two red lines represent (and are drawn slightly offset from)  the new, large-scale \hi\ tails T$_\mathrm{S}$ (south) and T$_\mathrm{N}$ (north) in the channels where we detect them (see Sect. \ref{sec:results}).}
\label{fig:cube}
\end{figure*}

\begin{figure*}
\centering
\includegraphics[width=18cm]{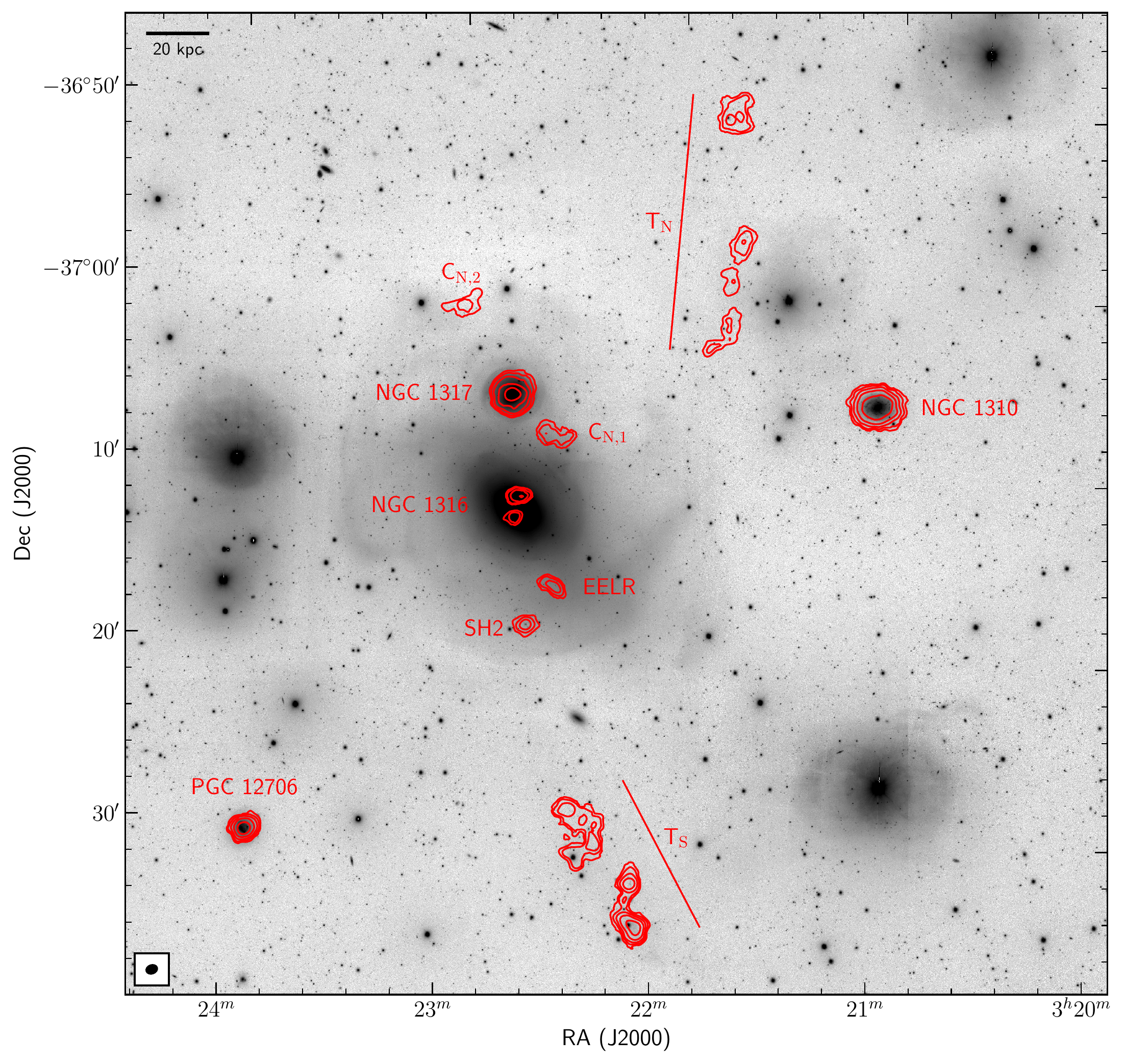}
\caption{Primary-beam corrected, constant column density \hi\ contours overlaid on a $r$-band image from the Fornax Deep Survey  \citep{iodice2017}  . The contour levels are $2.7\times10^{19} \times 2^n$ cm$^{-2}$ ($n=0,1,2,...$). The lowest contour corresponds to a $3\sigma$ signal in a single channel at the centre of the \hi\ cube (coincident with NGC~1316) but has lower significance further out. The Gaussian restoring PSF in the bottom-left corner has a FWHM of 36.7\arcsec\ $\times$ 28.1\arcsec\ and position angle 109 deg. The scale bar in the top-left corner represents 20 kpc ($\sim3.5$\arcmin\ at a distance of 20 Mpc). The \hi\ properties of the labelled sources are listed in Table \ref{table:mass}.}
\label{fig:overlay}
\end{figure*}

\begin{figure*}
\centering
\includegraphics[width=18cm]{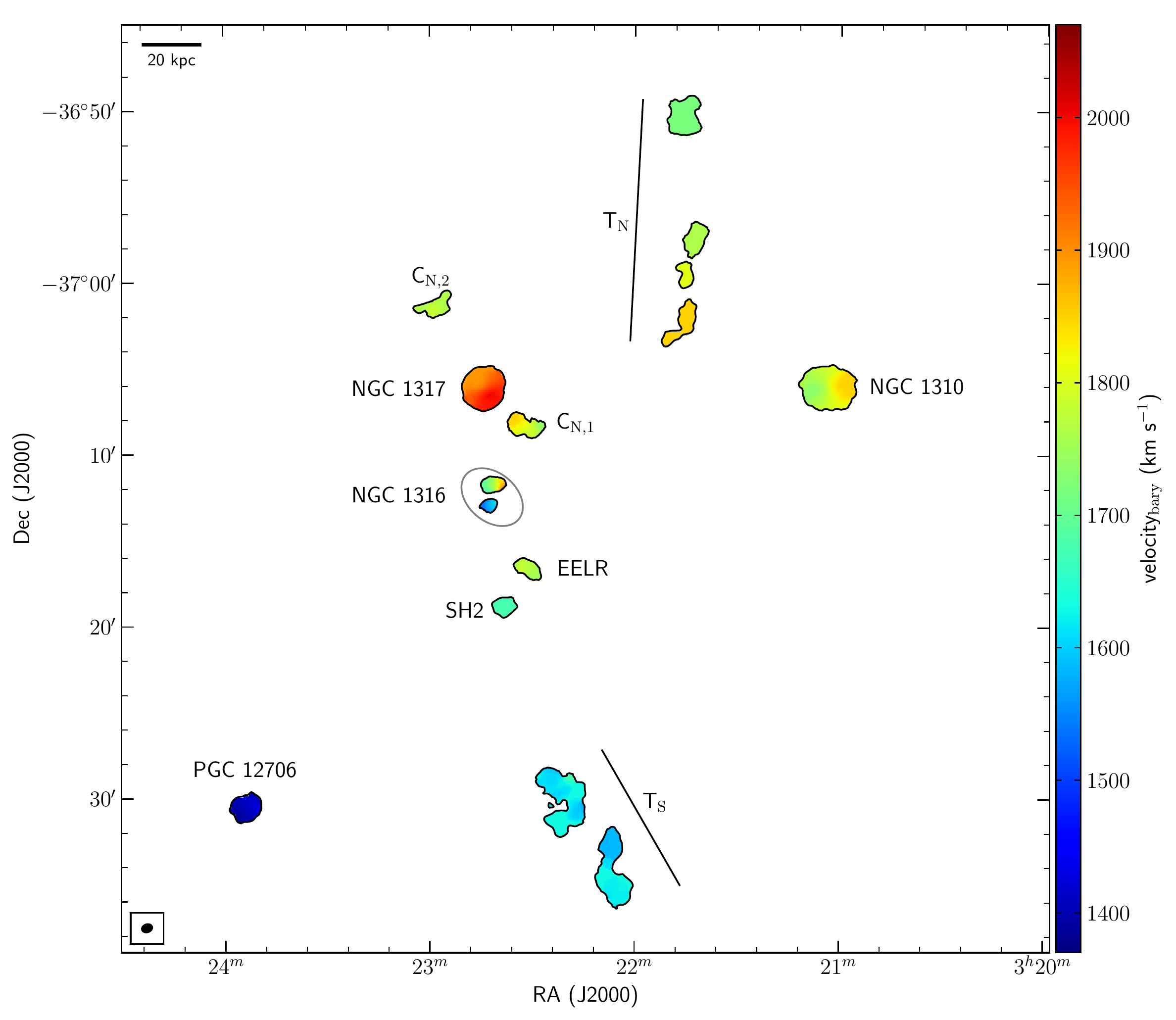}
\caption{\hi\ velocity field. Velocities are in the barycentric reference frame (optical definition). The velocity colour bar is centred at the systemic velocity of NGC~1316, 1760 \kms. The black \hi\ contour is the same as the first contour in Fig. \ref{fig:overlay} and has a column density of $2.7\times10^{19}$ cm$^{-2}$. The grey, open ellipse represents the stellar body of NGC~1316. The Gaussian restoring PSF in the bottom-left corner has a FWHM of 36.7\arcsec\ $\times$ 28.1\arcsec\ and position angle 109 deg. The scale bar in the top-left corner represents 20 kpc ($\sim3.5$\arcmin\ at a distance of 20 Mpc). The \hi\ properties of the labelled sources are listed in Table \ref{table:mass}.}
\label{fig:vfield}
\end{figure*}

As part of our data reduction we produced a radio continuum image of the field. The image includes the bright radio lobes of NGC~1316 \citep{ekers1983,fomalont1989}, whose flux density is $\sim100$ Jy in our data (mostly in the short baselines). The quality of the \hi\ cube is good despite this strong, diffuse radio emission. We present the radio continuum image in a separate paper.

\section{Results: \HIbf\ within and around NGC~1316}
\label{sec:results}

Figure \ref{fig:cube} shows the \hi\ cube channels in the velocity range where we detect emission: 1370 -- 2040 \kms\ (for reference, the barycentric systemic velocity of NGC~1316 is 1760 \kms  ; \citealt{longhetti1998} ). Several \hi\ detections already discussed by \cite{horellou2001} are visible. These include the resolved galaxies NGC~1310 and NGC~1317 (red ``x'' markers in Fig. \ref{fig:cube}) and the following \hi\ clouds within 5\arcmin\ of NGC~1316 ($\sim30$ kpc in projection; we mark this galaxy with a red ``+'' in the image):

\begin{itemize}
\item the extended emission-line region (EELR) south-west of NGC~1316 originally discovered by \cite{mackie1998}, which we detect at velocities 1730 -- 1820 \kms;
\item the star cluster complex SH2 south of NGC~1316 \citep{schweizer1980,richtler2012,richtler2017} at 1680 \kms;
\item and an \hi\ cloud just north of NGC~1316 at 1770 -- 1860 \kms\ (hereafter, C$_\mathrm{N,1}$).
\end{itemize}

\noindent  The following new \hi\ detections are also visible in Fig. \ref{fig:cube}:

\begin{itemize}
\item \hi\ in the central arcmin ($\sim5$ kpc) of NGC~1316 at velocities 1500 -- 1860 \kms;
\item a few \hi\ clouds forming a tail between 16\arcmin\ and 26\arcmin\ south of NGC~1316 (90 and 150 kpc in projection, respectively) at velocities 1590 -- 1640 \kms\ (hereafter T$_\mathrm{S}$);
\item a few \hi\ clouds forming a tail between 12\arcmin\ and 26\arcmin\ north-west of NGC~1316 (70 and 150 kpc in projection, respectively) at velocities increasing monotonically from 1730 to 1860 \kms\ when moving from north to south (hereafter T$_\mathrm{N}$);
\item an \hi\ cloud 12\arcmin\ north of NGC~1316 at 1770 \kms\ (70 kpc in projection; hereafter C$_\mathrm{N,2}$);
\item and, in the south-east corner, the galaxy PGC~12706 at 1370 -- 1460 \kms.
\end{itemize}

\noindent Most new \hi\ clouds are unresolved in velocity. Future data with better frequency resolution will allow us to better characterise their morphology and kinematics.

Figure \ref{fig:overlay} shows primary-beam-corrected, constant-column-density \hi\ contours overlaid on an $r$-band optical image from the Fornax Deep Survey  \citep{iodice2016,iodice2017,venhola2018} . Figure \ref{fig:vfield} shows the \hi\ velocity field. We label individual detections or systems of \hi\ clouds in both figures. We give their primary beam-corrected \hi\ flux and mass, with statistical errors, in Table \ref{table:mass}. The Table also lists the velocity range of each detection.

Our \hi\ flux values for NGC~1310, NGC~1317, EELR, SH2 and C$_\mathrm{N,1}$ agree with those published by \cite{horellou2001} within the errors. We do not confirm the detection of the \hi\ cloud $\sim5$\arcmin\ west of NGC~1316 shown in their Figure 2. If real, that $\sim 10^7$ \msun\ unresolved cloud would have been detected at a $\sim 10\  \sigma$ level in a single channel of our \hi\ cube. Conversely, \hi\ in the centre of NGC~1316 is too faint to be detected in the \cite{horellou2001} data ($\lesssim 1.5\sigma$ per channel); and the new \hi\ clouds at a large distance from NGC~1316 are relatively bright in our cube (especially those  in T$_\mathrm{S}$) but could not be detected in their data because of the higher noise and smaller primary beam.  This demonstrates the importance of the large primary beam of MeerKAT --- combined with the excellent sensitivity to diffuse emission --- for the study of nearby galaxies.

Altogether, we find $4.3\times10^7$ \msun\ of \hi\ in the centre of NGC~1316, another $9.6\times10^7$ \msun\ in the three already known clouds within $\sim30$ kpc from the galaxy (EELR, SH2 and C$_\mathrm{N,1}$) and, finally, a total of $5.8\times10^8$ \msun\ at larger distance,  70 to 150 kpc in projection (T$_\mathrm{N}$, T$_\mathrm{S}$ and C$_\mathrm{N,2}$). In total this amounts to $7.2\times10^8$ \msun\ of \hi\ associated with NGC~1316 on various scales, 14 times more than previously detected. At all distances from the centre of NGC~1316, gas to the north is redshifted relative to the systemic velocity, gas to the south blueshifted. In Sections \ref{sec:centre} to \ref{sec:tails} we discuss the properties of these  \hi\ components and their relation to the distribution of stellar light. In Sect. \ref{sec:other} we discuss the other galaxies detected in \hi\ in this field. In Sect. \ref{sec:discussion} we revisit the assembly history of NGC~1316 in view of the new detections.

\subsection{\HIsl\ in the centre of NGC~1316}
\label{sec:centre}

   \begin{table}
   {\centering
      \caption[]{\hi\ properties of the detected sources.}
         \label{table:mass}
         \begin{tabular}{lrrr}
            \hline
            \hline
            \noalign{\smallskip}
            Source      &  $F_\mathrm{HI}$ & \mhi & vel. range\\
                             &  (Jy \kms) & ($10^7$ \msun) & (\kms)  \\
            \noalign{\smallskip}
            \hline
            \noalign{\smallskip}
            NGC~1316                & $0.45\pm 0.04$   & $4.3\pm 0.4$ & 1500 -- 1860     \\
            EELR                         & $0.33\pm 0.03$   & $3.1\pm 0.3$ & 1780 -- 1820     \\
            SH2                           & $0.30\pm 0.02$   & $2.8\pm 0.2$ & 1680                 \\
            $\mathrm{C_{N,1}}$  & $0.39\pm 0.04$   & $3.7\pm 0.3$ & 1770 -- 1860      \\
            $\mathrm{C_{N,2}}$  & $0.29\pm 0.03$   & $2.7\pm 0.3$ & 1770                 \\
            $\mathrm{T_N}$        & $1.56\pm 0.08$   & $14.7\pm 0.8$  & 1780 -- 1860      \\
            $\mathrm{T_S}$        & $4.29\pm 0.10$   & $40.4\pm 1.0$  & 1590 -- 1640       \\
            NGC~1310                & $5.09\pm 0.09$   & $48.0\pm 0.9$  & 1780 -- 1910       \\
            NGC~1317                & $2.80\pm 0.06$   & $26.4\pm 0.6$  & 1860 -- 2040       \\
            PGC~12706              & $1.46\pm 0.06$   & $13.7\pm 0.6$  & 1370 -- 1460       \\
            \noalign{\smallskip}
            \hline
            \noalign{\smallskip}
         \end{tabular}
         
   \it Notes. \rm The \hi\ flux $F_\mathrm{HI}$ is derived from the primary-beam-corrected image. The \hi\ mass \mhi\ is calculated from $F_\mathrm{HI}$ using the equations in \cite{meyer2017} assuming a distance of 20 Mpc. The statistical errors are obtained propagating the noise of the primary-beam-corrected \hi\ cube to the 3D regions within which we integrate the flux. Velocities are in the barycentric reference frame (optical definition).}
   \end{table}

Our detection of \hi\ in the central $\sim5$ kpc of NGC~1316 adds one more component to the complex interstellar medium in this region: dust is clearly seen both in absorption \citep{carlqvist2010} and emission \citep{lanz2010}; molecular gas is also present and tracks the distribution of dust  \citep{horellou2001,morokumamatsui2019}; ionised gas appears to be distributed in a disc or ring with a polar orientation relative to the stellar rotation  \citep{schweizer1980,morokumamatsui2019}; and we have now finally detected \hi.

The \hi\ non-detection by \cite{horellou2001} demonstrated that atomic gas is not present in large amounts in this central region, resulting in a limit on the molecular-to-atomic hydrogen mass fraction \mhtwo/\mhi\ $\geq 6$. We now detect $4.3\times10^7$ \msun\ of \hi, which sets \mhtwo/\mhi\ $= 14$. Values of this order are not unusual in the centre of early-type galaxies \citep{oosterloo2010,serra2012a,young2014}.

To first order, the newly detected \hi\ tracks the distribution of dust and molecular gas. Its kinematics is also broadly consistent with that of molecular gas   \citep{horellou2001,morokumamatsui2019}  and ionised gas  \citep{schweizer1980,morokumamatsui2019}: the southern gas is blueshifted relative to the systemic velocity; the northern gas is distributed on an arc whose  velocity is redshifted relative to systemic velocity on the west side, and approaches systemic velocity when moving east  (Fig. \ref{fig:vfield}). We will present a detailed analysis of the distribution and kinematics of the multi-phase interstellar medium in the centre of NGC~1316 in a future paper.

\bigskip

\subsection{\HIsl\ at the outskirts of NGC~1316: EELR, SH2 and C$_\mathrm{N,1}$}

The better sensitivity of our data allows us to obtain new details on the \hi\ properties of EELR, SH2 and C$_\mathrm{N,1}$ compared to \cite{horellou2001}.  We can now measure their \hi\ mass more precisely, and resolve the gas distribution of EELR and C$_\mathrm{N,1}$.

In the EELR we detect $3.1\times10^7$ \msun\ of \hi. The emission is resolved and elongated with position angle $\sim45$ deg. This is a good match to the distribution of ionised gas shown by \cite{mackie1998}. The \hi\ column density is below $10^{20}$ cm$^{-2}$ across the entire cloud at the resolution of our data. We detect \hi\ in 3 velocity channels of our cube (1730 -- 1820 \kms; Fig. \ref{fig:cube}) but find no clear velocity gradient (Fig. \ref{fig:vfield}). We will present a multi-wavelength analysis of the EELR in a future paper.

The \hi-bright region SH2, which consists of a ring-shaped collection of 100-Myr-old star clusters \citep[\mstar\ $\sim 10^6$ \msun;][]{richtler2012,richtler2017}, remains unresolved in our \hi\ data both in velocity (at $\sim1680$ \kms) and on the sky. The angular and velocity resolution of our data is insufficient to search for any low-column-density \hi\ extension in the direction of an associated, intermediate-age globular cluster ($\sim15$\arcsec\ to the south-west and $\sim30$ \kms\ offset in velocity; \citealt{richtler2017}). The \hi\ mass of SH2 is $2.8\times 10^7$ \msun, which dominates over the stellar mass by approximately an order of magnitude.

Finally, we find $3.7\times10^7$ \msun\ of \hi\ in C$_\mathrm{N,1}$. This cloud is located at the northern edge of the stellar body of NGC~1316 and is elongated in a direction approximately parallel to the optical isophotes in that region. The \hi\ column density is below $10^{20}$ cm$^{-2}$ at the resolution of our data. Gas in this cloud is visible in 3 velocity channels of our cube (1770 -- 1860 \kms; Fig \ref{fig:cube}), and we observe a gradient with velocity increasing from west towards east (Fig. \ref{fig:vfield}).

\subsection{\HIsl\ at a large distance from NGC~1316: T$_\mathrm{N}$, T$_\mathrm{S}$ and C$_\mathrm{N,2}$}
\label{sec:tails}

The vast majority of the \hi\ associated with NGC~1316 is distributed at large  radii  in the two tail-like complexes T$_\mathrm{N}$ and T$_\mathrm{S}$. At the resolution of our data T$_\mathrm{N}$ includes 4 or 5 \hi\ clumps (Fig. \ref{fig:overlay}), which  form a coherent 3D structure in the \hi\ cube  from 70 to 150 kpc (projected) north-west of NGC~1316 (Figs. \ref{fig:cube} and \ref{fig:vfield}). The southernmost clump is redshifted by $\sim100$ \kms\ relative to systemic velocity. The \hi\ velocity then decreases progressively as we move north, and the northernmost clump is blueshifted by $\sim30$ \kms\ relative to systemic velocity. The clumps could simply be the densest parts of an underlying, diffuse distribution of gas stretching approximately north to south. The \hi\ distribution appears long ($\sim80$ kpc in projection) and narrow ($\lesssim6$ kpc in projection; it is mostly unresolved by the \hi\ PSF along its transverse direction). The \hi\ column density remains low, $\lesssim 10^{20}$ cm$^{-2}$, across the entire extent of T$_\mathrm{N}$ at the resolution of our data.

The structure of the more massive T$_\mathrm{S}$ might not be as simple as that of T$_\mathrm{N}$. The \hi\ is distributed in two main components, the southernmost of which includes in turn two individual, bright \hi\ clumps (Fig. \ref{fig:overlay}).  All of the gas in T$_\mathrm{S}$ is detected in just two channels of our \hi\ cube, both blueshifted relative to systemic. The emission exhibits no clear  ordered 3D structure in the \hi\ cube  at the velocity resolution of our data (Figs. \ref{fig:cube} and \ref{fig:vfield}). Gas in T$_\mathrm{S}$ reaches a relatively high column density, nearly $10^{21}$ cm$^{-2}$ in the two southernmost clumps. Higher angular resolution data would likely bring the peak column density to even higher values.

Finally, C$_\mathrm{N,2}$ is the only isolated \hi\ cloud detected at a large distance from NGC~1316. It is faint and nearly unresolved. It is located $\sim70$ kpc north of the galaxy in projection and, like T$_\mathrm{N}$, it is redshifted relative to systemic. This cloud makes up just 5 percent of all the \hi\ detected at a large distance from NGC~1316.

\begin{figure}
\centering
\includegraphics[width=9cm]{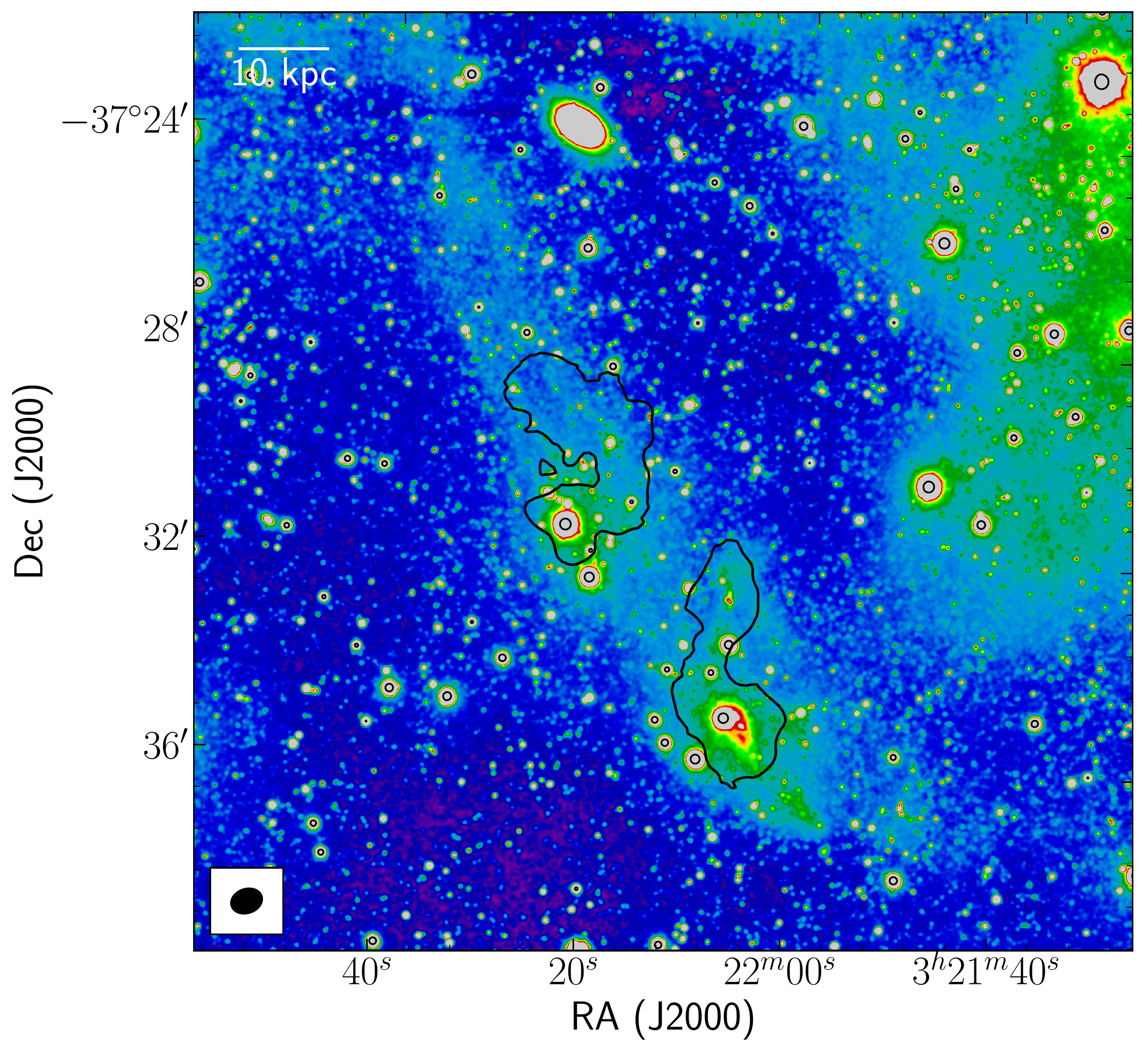}
\caption{ Relation between the \hi\ tail T$_\mathrm{S}$ and the optical tidal feature L5. The image shows a single \hi\ contour at a column density of $2.7\times10^{19}$ cm$^{-2}$ (black) overlaid on a stack of the $g$-, $r$- and $i$-band images from the Fornax Deep Survey  \citep{iodice2017}  shown in false colour with a linear stretch. The optical image was convolved with a 4\arcsec\ Gaussian (FWHM). The black circles indicate bright foreground stars. The \hi\ Gaussian restoring PSF in the bottom-left corner has a FWHM of 36.7\arcsec\ $\times$ 28.1\arcsec\ and position angle 109 deg. The scale bar in the top-left corner represents 10 kpc ($\sim100$\arcsec\ at a distance of 20 Mpc).}
\label{fig:L5}
\end{figure}

These new \hi\ detections at large  radii  are associated with the fine structure identified through deep optical imaging by \cite{schweizer1980} and \cite{iodice2017}\footnote{See also https://images.datacentral.org.au/malin/DEEP/060 .}. The prime example is T$_\mathrm{S}$, which is associated with the stellar feature L5. In particular, the overlay of \hi\ contours on top of a \it gri \rm stacked image (Fig. \ref{fig:L5}) shows that the northern, more diffuse part of T$_\mathrm{S}$ is associated with a faint, diffuse portion of L5; while the two southernmost \hi\ clumps in T$_\mathrm{S}$ correspond to a section of L5 which is clumpier and brighter than the rest of the feature. The $r$-band surface brightness of this bright, southern region reaches 25 mag arcsec$^{-2}$ \it vs. \rm the $>26$ mag arcsec$^{-2}$ more typical of L5 as a whole. In the same region we find a high \hi\ column density  and, based on a preliminary data reduction of narrow-band imaging obtained recently by our group with the ESO/VST telescope, clumps of H$\alpha$+\Nii\ emission. It is therefore likely that some star formation is on going --- or has occurred recently --- within them.

Figure \ref{fig:L5} also shows that the association between \hi\ and optical light does not hold across the entire L5. First, a bright clump of stellar light is visible 1\arcmin\ -- 2\arcmin\ south-west of T$_\mathrm{S}$, in a region where we do not detect any \hi\ at the sensitivity and resolution of our MeerKAT data  (see Table \ref{table:obs}) . This region is not detected in H$\alpha$+\Nii\ either. Furthermore, fainter, diffuse light belonging to L5 is clearly visible for several arcmin north-east of T$_\mathrm{S}$. Here, too, we find no \hi.

The other \hi\ detections T$_\mathrm{N}$ and C$_\mathrm{N,2}$ are also associated with known optical features, although more ambiguously so than T$_\mathrm{S}$ with L5. First, the southern end of T$_\mathrm{N}$ coincides with the south-west end of the L9 feature discovered by \cite{iodice2017}. However, while T$_\mathrm{N}$ continues northwards to a region with no optical emission, L9 bends towards east. Similarly, C$_\mathrm{N,2}$ is located within another optical feature identified by \cite{iodice2017}, L7. However, the optical surface brightness of this feature does not change significantly at the location of the \hi\ cloud.

\subsection{\HIsl\ in galaxies near NGC~1316}
\label{sec:other}

Besides NGC~1316, the observed field includes three relatively large, \hi-rich galaxies: NGC~1317, NGC~1310 and PGC~12706 (in order of increasing projected distance from NGC~1316; see Fig. \ref{fig:overlay}). These have a $+200$, $\sim0$ and $-350$ \kms\ offset relative to the recessional velocity of NGC~1316, respectively. The first two are likely part of the NGC~1316 group. NGC~1310 is slightly in the foreground  \citep{fomalont1989,anderson2018}, while the exact 3D location of NGC~1317 is more uncertain. The case of PGC~12706 is less clear as the velocity difference from NGC~1316 is relatively large. This galaxy could be a few Mpc in front of the group.

Our data do not reveal any striking new \hi\ features in NGC~1317 compared with the data of \cite{horellou2001}. Despite our slightly more sensitive data, we confirm their conclusion that the \hi\ disc is confined to within the stellar disc and that the galaxy is \hi-deficient, indicating that some gas has likely been removed from its outskirts possibly following the interaction with the intra-group medium. We also confirm the lack of any asymmetries or tidal features in the \hi\ disc. It is therefore likely that the 3D distance between NGC~1317 and the $\sim10$ times more massive NGC~1316 is too large for  NGC~1317 to be tidally perturbed.

Our data for NGC~1310 are significantly more sensitive than those of \cite{horellou2001} because of our larger primary beam (the galaxy is $\sim20$\arcmin\ from the pointing centre). Unlike in that previous study, we find the \hi\ column density to be symmetric across the disc. Therefore, this galaxy, too, shows no signs of interaction with NGC~1316. At the resolution of our data there appears to be an \hi\ depression in the galaxy centre. Furthermore, the \hi\ disc is more extended than the stellar disc, which is common for galaxies of this late morphological type.

Finally, we detect for the first time \hi\ in PGC~12706. This galaxy has a peculiar optical morphology characterised by a bright, distorted ring of star formation, likely originating from a collision \citep{madore2009}. The \hi\ is unresolved in our data and, therefore, gives us no clue about the origin of the ring. The \hi\ is very bright, though, and future observations with higher angular and velocity resolution will be useful to understand the nature of this object.

\section{The assembly history of NGC~1316}
\label{sec:discussion}

Our detection of $7\times 10^8$ \msun\ of \hi\ within and around NGC~1316 --- mostly at large  radii  and partly associated with optical tidal features --- provides new constraints on the assembly history of this galaxy. In particular, it gives renewed support to the hypothesis that NGC~1316 formed  through a merger between a dominant, gas-poor early-type galaxy and a smaller, gas-rich spiral.

As discussed in \cite{schweizer1980} and \cite{iodice2017}, NGC~1316 has likely experienced multiple mergers with mass ratio $\lesssim$ 10:1 in the last few billion years. The picture emerging from those studies is that the oldest merger is likely traced by the longest and faintest tidal features --- L5 to the south, shown in Fig. \ref{fig:L5}, and L9 to the north --- and has occurred at least 1 Gyr ago. Subsequent mergers might be responsible for the smaller and brighter tidal features visible in Fig. \ref{fig:overlay}, such as those to the east (known as L2) and west (L1). Our detection of \hi\ exactly coincident with L5 and in the proximity of L9 establishes an important new fact: the oldest merger was gas-rich. The lack of \hi\ at the location of the brighter, more recent tidal features suggests that, in comparison, subsequent mergers were gas-poor.

This result suggests a possible causal relation between that first, gas-rich merger and other evidence that cold gas has flowed recently into the centre of NGC~1316, where it has been partly consumed by star formation. The galaxy hosts a multi-phase interstellar medium dominated by molecular gas and kinematically decoupled from the stars in the central few kiloparsecs of the stellar body (\citealt{schweizer1980,horellou2001,morokumamatsui2019}; this work). Furthermore, optical and infrared spectroscopy reveal intermediate-age stellar populations, which is best interpreted in terms of a $\sim10$ percent (in mass) of $\sim 1$-Gyr-old stars on top of an older population \citep{kuntschner2000,silva2008}.  And, finally, the timescale of this star formation event is comparable to that for the formation of a relatively young (1 --3 Gyr old) population of globular clusters, which indeed \cite{goudfrooij2001} and \cite{sesto2018} used to date the initial merger.

We therefore propose that the same gas-rich merger that produced stellar and gaseous tidal features at large radii is also responsible for the accretion of gas into the central regions of NGC~1316. During mergers, non-axisymmetries generated by tidal forces can drive some of the gas inward, where it triggers a starburst, while the same tidal forces can pull stars and the remaining gas out to large radii, where they can remain visible for billions of years \citep[e.g.,][]{dimatteo2007,dimatteo2008}. In other known merger remnants, such as those discussed in Sect. \ref{sec:intro}, cold gas at large radius is mostly atomic while, in the central region, most of it is in molecular form and associated with recent star formation. The same is true in NGC~1316, where we find identical masses of neutral hydrogen in the centre (mostly molecular) and at large radii \citep[mostly atomic;][]{horellou2001,morokumamatsui2019}.

The association between stellar light and \hi\ in merger remnants is not always tight. Gas-poor optical features and ``dark'' \hi\ features exist in many systems like those mentioned in Sect. \ref{sec:intro}. In this sense, the lack of stellar light across most of the \hi\ tail T$_\mathrm{N}$ and the lack of \hi\ across, e.g., most of L9 are not too surprising (see Sect. \ref{sec:tails}). The definite association between T$_\mathrm{S}$ and L5 is more consequential in the discussion about the origin of NGC~1316 than the lack of association between stellar light and \hi\ at other locations.

The above, qualitative discussion shows that, based on the new \hi\ data and on the large body of previous observational results, many features of NGC~1316 can be traced back to a gas-rich merger occurred at most a few billion yers ago --- likely followed by subsequent mergers involving smaller amounts of cold gas. We supplement that discussion with some more quantitative arguments on the nature of that initial merger. As a first step we revisit  \cite{lanz2010}'s calculation of the expected \hi\ mass of NGC~1316 in the context of its merger origin. As discussed in Sect. \ref{sec:intro}, NGC~1316 hosts $>10$ times more dust than expected given the typical dust-to-stellar mass ratio of galaxies with its morphology and stellar mass \citep{smith2012}. We therefore assume that all of the dust currently present within NGC~1316 ($\sim2\times 10^7$ \msun;  \citealt{draine2007,lanz2010,galametz2012}) has an external origin and comes from an Sa progenitor that merged with an early-type galaxy\footnote{Our choice of progenitors aims at keeping the merger mass ratio below $\sim$ 10:1. Given the low cold gas-to-stellar mass ratio of early-type spirals, choosing an Sa galaxy as the smaller progenitor maximises its stellar mass and, therefore, minimises the merger mass ratio at fixed cold-gas mass. Choosing a later-type spiral progenitor would give a mass ratio too large to be consistent with previous results on the merger that generated the extended optical tidal features like L5 \citep{schweizer1980,mackie1998,goudfrooij2001}.}.  Galaxies of Sa type have a typical dust-to-stellar mass ratio of $\sim5\times 10^{-4}$ \citep{smith2012}. Therefore, the Sa's stellar mass  was likely to be  $\sim4\times 10^{10}$ \msun. The merger that formed NGC~1316, whose stellar mass is now $\sim6\times10^{11}$ \msun\ \citep{iodice2017},  appears then to have had  a mass ratio of 14:1. This would be sufficient to form the  long  tidal features visible in its optical images  such as L5 and L9.

How much \hi\ would the Sa progenitor bring into NGC~1316? Galaxies with this Hubble type and stellar mass have a stellar surface mass density $\mu_\star \sim 10^9$ \msun/kpc$^2$ (based on the sample of \citealt{nair2010}). The scaling relation between \mhi/\mstar, \mstar\ and  $\mu_\star$ \citep{brown2015} tells us that the galaxy  likely had  \mhi /\mstar\  $\sim0.05$. We thus expect the Sa progenitor of NGC~1316 to have had \mhi\ $\sim 2\times10^9$ \msun. This result is in good agreement with the earlier calculation by \cite{lanz2010}.

Our detection of $7\times 10^8$ \msun\ of \hi\ around NGC~1316 increases the \hi\ mass of this galaxy by a factor of 14 compared to previous work, bringing it within a factor of just $\sim3$ of the \hi\ mass expected for the progenitor Sa galaxy. This is now an acceptable difference considering: \it i) \rm the large scatter around many of the scaling relations used for the above calculation (e.g., scaling relations involving \mhi /\mstar\ have a typical scatter between 0.5 and 1 dex; see \citealt{catinella2018}); \it ii) \rm the uncertainty on some of the quantities that go into those scaling relations (in this case, dust mass, stellar mass, and stellar mass surface density);  \it iii) \rm the possibility that the Sa progenitor was \hi-deficient following the removal of some of its cold gas through an interaction with the intra-group medium (indeed, as discussed in Sect. \ref{sec:other}, the nearby galaxy NGC~1310 is \hi-deficient); \it iv) \rm the possibility that some \hi\ in the field is below our detection limit;  and \it v) \rm the fact that during a merger some \hi\ will inevitably be lost to star formation in the remnant  (as discussed above)  and to ionisation once tidal forces diffuse it into the inter-galactic medium.

The above calculation should not be over-intepreted. For example, it is possible that the spiral progenitor was a  slightly more massive galaxy with a   slightly lower \mhi /\mstar\ ratio. However, the same calculation suggests that a dramatically different merger is unlikely. A merger between two equal-mass galaxies --- of which at least one gas-rich --- is ruled out as it would involve a much larger gas mass than observed. An additional consistency argument is that galaxies similar to the hypothetical Sa progenitor discussed above are present in the immediate surroundings of NGC~1316, namely NGC~1310 and NGC~1317. It is possible that one such galaxy has merged with a larger early-type galaxy to form NGC~1316.

\begin{figure}
\centering
\includegraphics[width=9cm]{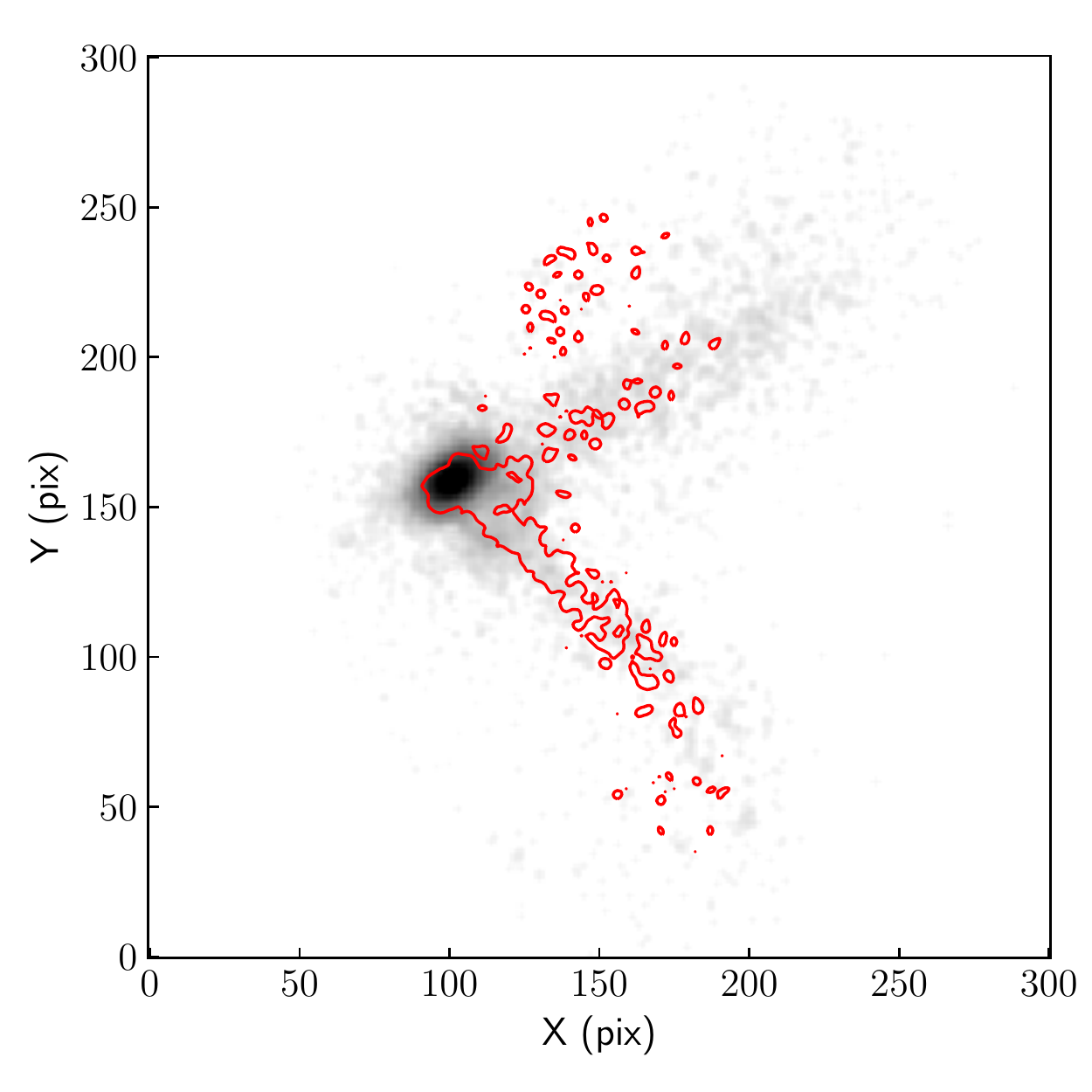}
\caption{Snapshot of GalMer merger \# 971  \citep{chilingarian2010}  taken 2.5 Gyr after the start of the simulation and $\sim2$ Gyr after the first pericentral passage. The background greyscale image shows the distribution of stellar mass on a logarithmic scale. The red contours show the distribution of gas. Both stellar and gas images have been smoothed for clarity. The image is $750\times 750$ kpc$^2$, but note that we ignore the physical scale of the system in our comparison to NGC~1316 for reasons explained in the text.}
\label{fig:galmer}
\end{figure}

We further investigate the consistency between our observational results and the expectation for a $\sim$ 10:1 early-type + spiral merger remnant using the GalMer database of simulated galaxy mergers \citep{chilingarian2010}. GalMer includes mergers over a range of galaxy morphologies, mass ratios and orbital parameters. We have inspected several S0+Sa 10:1 mergers and found that those on prograde orbits typically result in the presence of  low-surface-brightness  tidal material (both stars and gas) at large  radii  due the large total angular momentum of the system --- like in NGC~1316.

Given the limited size of the GalMer database it would be unreasonable to expect an exact match to the properties of NGC~1316. For example, the total stellar mass of the S0+Sa 10:1 mergers is $\sim7$ times smaller than that of NGC~1316, and the range of orbital parameters explored in the database is limited. Furthermore, the resolution of the GalMer simulations limits a thorough treatment of the complex gas physics, and no feedback from an active galactic nucleus is included. With these caveats in mind, we note however the following important consistencies between NGC~1316 and the GalMer merger \# 971 observed $\sim 2$ Gyr after the first pericentral passage (Fig. \ref{fig:galmer}):

\begin{itemize}
\item the merger exhibits multiple large-scale stellar tidal tails which include a gas counterpart for at least part of their extent; this agrees with our detection of \hi\ along part of L5 and at the start of L9; the tails are still expanding at the time of the snapshot in Fig. \ref{fig:galmer}; the exact shape and size of the tails varies with viewing angle (e.g., nearly radial in Fig. \ref{fig:galmer}, more curved/broad on other projections), but overall they are always clearly visible;
\item on some projections, the velocities of the two gas tails lay on opposite sides relative to the systemic velocity of the remnant, as in NGC~1316; and the velocity of the gas approaches the systemic velocity when moving towards larger radii along a tail, as in T$_\mathrm{N}$ (Fig. \ref{fig:vfield});
\item the remnant hosts approximately equal amounts of gas in the central region and in the outer tails; this is consistent with NGC~1316, where equal amounts of neutral hydrogen are found in the centre (mostly molecular) and in the tails (mostly atomic; GalMer does not distinguish between these two phases);
\item the stellar tidal tails are only $\sim 0.1$ mag bluer than the stellar body in $g-r$ colour, consistent with the idea that they formed several billion years ago, and with the small colour difference between most tidal features and the main stellar body in NGC~1316 \citep{iodice2017};
\item the stellar body rotates about its short axis as in NGC~1316 \citep{bosma1985,arnaboldi1998}.
\end{itemize}

A comparison between NGC~1316 and a set of GalMer mergers was presented also by \cite{iodice2017}. They found that the same 10:1 merger analysed here (\#971 in the GalMer database, which they also observed $\sim 2$ Gyr after the first pericentral passage) provides a good match to the structure of the stellar body of NGC~1316 and to the surface brightness of its outer tidal features L5 and L9 --- while a 1:1 merger between two disc galaxies does not. In that paper, the 10:1 merger is favoured also because it has no gaseous tails, consistent with the \hi\ results available at that time \citep{horellou2001}. We now detect extended \hi\ tails in NGC~1316 but, as argued above, this does not rule out a 10:1 merger. In fact, Fig. \ref{fig:galmer} shows that the 10:1 merger remnant, too, hosts extended gas tails. Compared to the analysis in \cite{iodice2017} the gas tails are now visible because we smooth the simulations, enhancing low-surface-brightness features, and we choose a projection which makes them stand out. This projection does not match the orientation of NGC~1316's stellar body (unlike in \citealt{iodice2017}) but this is not of real concern since there are no reasons to expect an exact match between the galaxy and the simulation.

There are also significant differences between GalMer merger \# 971 and NGC~1316.  For example, gas in its centre is kinematically aligned with the stars (unlike in NGC~1316; \citealt{schweizer1980,morokumamatsui2019}); and the length of its tidal tails exceeds the observed one. However, considering the limitations mentioned above and the complex recent history of NGC~1316, which includes episodes of nuclear activity \citep[e.g.,][]{ekers1983,fomalont1989} and further minor mergers \citep{schweizer1980,iodice2017}, this comparison should only be viewed as a consistency check. The main message is that some of the first-order features of NGC~1316 can be reproduced by an S0+Sa $\sim$ 10:1 prograde merger, which, as argued by \cite{lanz2010} and confirmed above in this section, is required to explain the overall composition of the galaxy's interstellar medium.

To conclude, it would be unreasonable to expect that a single merger can explain all the properties of NGC~1316. As thoroughly discussed by \cite{schweizer1980} and \cite{iodice2017}, NGC~1316 is likely continuing to accrete satellites. These might explain, for example, some of the minor \hi\ clouds in Fig. \ref{fig:overlay}. However, our results of a similar amount of neutral gas in the inner regions of NGC~1316 (mostly molecular) as in the tails at large  radii  (atomic) associated with optical tidal features, together with the evidence of central star formation in the last few Gyr, are consistent with the merger hypothesis. A quantitative comparison between NGC~1316 and simulations is challenging, but the picture emerging from this diverse set of observations is becoming increasingly clearer.

\section{Summary}

We have observed the nearby merger remnant NGC~1316 at the outskirts of the Fornax galaxy cluster with MeerKAT in search of \hi\ associated with its prominent stellar tidal features. We detect a total of $7\times10^8$ \msun\ of \hi\ within and around this galaxy, 14 times more than in previous observations. Most of the \hi\ is distributed in two tail-like complexes at large  radii, 70 -- 150 kpc in projection. The southern tail is associated with the optical tidal feature known as L5. It hosts clumps of dense \hi\  ($\sim10^{21}$ cm$^{-2}$)  and exhibits signs of recent star formation. The northern tail is only partly associated with a known optical feature and is characterised by a low \hi\ column density  ($\lesssim10^{20}$ cm$^{-2}$) . In addition to several scattered \hi\ clouds at various distance from NGC~1316, we also detect for the first time \hi\ in the centre of the galaxy: $4\times10^7$ \msun\ closely associated with dust and molecular gas in the central $\sim5$ kpc.

These results shed new light on the assembly history of NGC~1316. The total \hi\ mass and the location of most of the \hi\ at large  radii, near known stellar tidal features, together with the presence of a similar mass of gas in a molecular form in the galaxy centre, where a starburst has recently occurred, strengthen the hypothesis that NGC~1316 formed  through a gas-rich merger a few billion years ago, followed by subsequent, gas-poorer accretion of satellite galaxies. To first order, the properties of NGC~1316 are consistent with this initial merger having occurred between a dominant, gas-poor early-type galaxy and a $\sim10$ times smaller gas-rich spiral.

Future MeerKAT observations of this field conducted as part of the MeerKAT Fornax Survey \citep{serra2016b} will achieve significantly better sensitivity as well as angular and velocity resolution than those presented here. We will reach a similar \hi\ column density sensitivity  ($\sim3\times10^{19}$ cm$^{-2}$)  at $\sim3$ times better angular resolution  (10\arcsec)  and with $>40$ times narrower  velocity channels ($\sim1$ \kms) ; while  at the 30\arcsec\ angular resolution  of the current data we will detect \hi\ down to a column density of a few times $10^{18}$ cm$^{-2}$. The future data will thus unveil additional details of the distribution of \hi\ within and around NGC~1316.

\begin{acknowledgements}

 We thank Fernando Camilo for useful comments on an early draft of this paper. We are grateful to the full MeerKAT team at SARAO for their work on building and commissioning MeerKAT.  This project has received funding from the European Research Council (ERC) under the European Union's Horizon 2020 research and innovation programme (grant agreement no. 679627; project name FORNAX).  PK acknowledges being partially supported by the BMBF project 05A17PC2 for D-MeerKAT. The MeerKAT telescope is operated by the South African Radio Astronomy Observatory, which is a facility of the National Research Foundation, an agency of the Department of Science and Technology. 

\end{acknowledgements}

%
%

\bibliographystyle{aa} 
\bibliography{../../myrefs} 

\end{document}